\documentstyle[12pt]{article}
\begin{document}

\thispagestyle{empty} \centerline{\large\bf Manifestations the
hidden symmetry of Coulomb problem in the} \centerline{\large\bf
relativistic quantum mechanics - from Pauli to Dirac electron }
\bigskip
\begin{center}
Tamari T. Khachidze and Anzor A. Khelashvili\footnote{Email:
anzorkhelashvili@hotmail.com}\\ \vspace*{.5cm} {\em 3 Chavchavadze
Avenue, Department of Theoretical Physics,
\\Iv. Javakhishvili
Tbilisi State University, Tbilisi 0128, Georgia}\\ \vspace*{.5cm}
\end{center}
\vskip.5cm

\begin{abstract}
The theorem known from Pauli equation about operators that
anticommute with Dirac's $K$-operator  is generalized to the Dirac
equation. By means of this theorem the operator is constructed
which governs the hidden symmetry in relativistic Coulomb problem
(Dirac equation). It is proved that this operator coincides with
the familiar Johnson-Lippmann one and is intimately connected to
the famous Laplace-Runge-Lenz (LRL) vector. Our derivation is very
simple and informative. It does not require a longtime and tedious
calculations, as is offten underlined in most papers.
\end{abstract}
\vspace*{.3cm}

\vspace*{1.0cm} PACS :  \\

\vspace*{0.3cm} Key words: :  Coulomb potential, hidden symmetry,
accidental degeneracy, Witten superalgebra and supercharges, Lamb
shift.
\\

\newpage

It is well known that the Coulomb Problem has additional dynamical
symmetry in classical mechanics as well as in non-relativistic
quantum mechanics. This symmetry guarantees the classical motion
on the closed orbits. At the same time, it is responsible to the
"accidental" degeneracy of hydrogen atom spectrum. The nature of
this phenomena was explained by V.Fock \cite{Fock}, V. Bargman
\cite{Barg}, W.Pauli \cite{Pauli} and others many  years ago.

There are no closed orbits in relativistic Kepler problem,
however. Recall that Sommerfeld \cite{Somer} obtained his famous
energy levels of the hydrogen atom by transforming to a rotating
coordinate system in which the relativistic precession (the
familiar "rosette" motion) was eliminated and closed orbits occur.
This means that there is still some residual hidden symmetry of
the Kepler problem in relativistic mechanics as well.

This consideration is enhanced by recent investigations. It was
shown \cite{KA}, that the Dirac equation still have some "hidden"
symmetry, which ultimately provides a certain algebra (so called,
Witten algebra), that gives a quantum-mechanical supersymmetry of
this problem.

In other words, the additional symmetry (conservation of the
Laplace-Runge-Lenz, hereafter, LRL vector) transforms into a
certain supersymmetry in microworld (hydrogen atom), which
controls the degeneracy  of the hydrogen spectra in the Dirac
equation.

Together with bosonic degrees of freedom supersymmetry requires
existence of fermionic degrees of freedom as well. In this purpose
it is easier to discuss the motion in a Coulomb field of a Pauli
particle (nonrelativistic spin-$1/2$ particle with kinematically
independent spin). This simplicity results from using the spin
operator $\vec{\sigma}$, together with the two vector operators
$\vec{l}$ (angular momentum) and $\vec{A}$ (LRL vector). With the
help of spin operator the Dirac non-relativistic operator's
analogue is constructed, $K_p=-(2\vec{\sigma}\vec{l}+1)$. It
commutes with the Pauli Hamiltonian and allows introduction of
$Z_2$ graduation in the Hilbert space by making use of parity
operator $P_k=K_p/|k|$ and the physical states (operators) are
divided into even and odd ones.

To find  odd operators certain theorem is used \cite{BL},  which
now we will generalize below for the Dirac case.

The Dirac Hamiltonian in Coulomb potential has the form
\begin{equation}\label{Dirham}H=\vec{\alpha}\vec{p}+\beta m-{a\over r}~,~~~a\equiv Ze^2=Z\alpha~.\end{equation}

It is known, that the total momentum operator
$\vec{J}=\vec{l}+\vec{\Sigma}/2$ commutes with $H$, where
$\vec{\Sigma}$ is the spin matrix
\[\vec{\Sigma}=\gamma^5\vec{\alpha}=
\left(\begin{array}{cc}\vec{\sigma} & 0\\0 &
\vec{\sigma}\end{array}\right)~.\]

It is connected to the usual Dirac's matrices by relation
$\vec{\Sigma}=\gamma^5\vec{\alpha}$. It is the Dirac's $K$
operator
\begin{equation}\label{Dirope}K=\beta\left(\vec{\Sigma}\vec{l}+1\right)~,\end{equation}
which  commutes with $H$. It has the following properties
\begin{equation}\beta K=K\beta~,~~~~~\gamma^5 K=-K\gamma^5~.\end{equation}

The spectrum of Hamiltonian (\ref{Dirham}) is degenerate with
respect to two signs of the $K$'s eigenvalues, $\pm k$ \cite{BL}.
It is evident that the operator, that interchange signs of the
$k$, must be anticommuting with $K$. If such anticommuting
operator at the same time would be commuting with Hamiltonian,
there arises the additional symmetry, which is exactly Witten's
superalgebra.

Let us generalize the above mentioned theorem from Pauli to Dirac
electron. \newline{\bf Theorem:}

Suppose $\vec{V}$ is a vector with respect to the orbital angular
momentum $\vec{l}$ that is also perpendicular to $\vec{l}$, i.e.,
\[\vec{l}\times\vec{V}+\vec{V}\times\vec{l}=2i\vec{V}~,~~~~\vec{l}\vec{V}=\vec{V}\vec{l}=0~.\]
then $K$ anticommutes  with the $\vec{J}$ scalar,
$\vec{\Sigma}\vec{V}$.
\newline{\bf Proof:}
Let us consider a product
$(\vec{\Sigma}\vec{l})(\vec{\Sigma}\vec{V})$. Exploiting the
properties of Dirac matrices, one can establish that
\[(\vec{\Sigma}\vec{l})(\vec{\Sigma}\vec{V})=(\vec{l}\vec{V})-i(\vec{\Sigma},~\vec{l}\times \vec{V})
=i(\vec{\Sigma},~2i\vec{V}-\vec{V}\times\vec{l})=-2\vec{\Sigma}\vec{V}-i\vec{\Sigma}(\vec{V}\times\vec{l})~.\]

Therefore

\begin{equation}\label{relat}\left(\vec{\Sigma}\,\vec{l}+1\right)\left(\vec{\Sigma}\,\vec{V}\right)=-\vec{\Sigma}\,\vec{V}-i\vec{\Sigma}(\vec{V}\times\vec{l})
~.\end{equation}

Now consider the same product in reversed order

\[(\vec{\Sigma}\vec{V})(\vec{\Sigma}\vec{l})=\vec{V}\vec{l}+i\vec{\Sigma}(\vec{V}\times\vec{l})=i\vec{\Sigma}(\vec{V}\times\vec{l})~.\]

Hence

\[\vec{\Sigma}\vec{V}(\vec{\Sigma}\vec{l}+1)=\vec{\Sigma}\vec{V}+(\vec{\Sigma}\vec{V})(\vec{\Sigma}\vec{l})=\vec{\Sigma}\vec{V}+i\vec{\Sigma}(\vec{V}\times\vec{l})
=-(\vec{\Sigma}\vec{l}+1)(\vec{\Sigma}\vec{V})~.\]

In the last step we made use of equation (\ref{relat}). Therefore
we obtain

\begin{equation}\{\vec{\Sigma}\vec{l}+1,~\vec{\Sigma}\vec{V}\}=0~.\end{equation} Now it
follows finally, that
\begin{equation}K(\vec{\Sigma}\vec{V})=-(\vec{\Sigma}\vec{V})K~.\end{equation}

It is evident that the class of anticommuting with $K$ (or,
$K$-odd) operators is not confined by these operators only - any
operator of type $\hat{O}(\vec{\Sigma}\vec{V})$, where $\hat{O}$
is commuting with $K$, but otherwise arbitrary, also is $K$-odd.

Let mention, that in the framework of constraints of above
theorem, the following very useful relation takes place
\begin{equation}\label{usrel}K\left(\vec{\Sigma}\vec{V}\right)=
-i\beta\left(\vec{\Sigma}~,~{1\over
2}\left[\vec{V}\times\vec{l}-\vec{l}\times\vec{V}\right]\right)~.
\end{equation}

One can see that the antisymmetrized vector product, familiar to
LRL vector appears on the right-hand-side of this relation.

Important special cases, resulting from the above theorem include
$\vec{V}=\hat{\vec{r}}$ (unit radial vector), $\vec{V}=\vec{p}$
(linear momentum) and $\vec{V}=\vec{A}$ (LRL vector), which has
the following form
\begin{equation}\vec{A}=\hat{\vec{r}}-{i\over 2ma}\left[\vec{p}\times\vec{l}-\vec{l}\times\vec{p}\right]~.\end{equation}

According to (\ref{usrel}), there appears one relation between
these three odd operators
\begin{equation}\vec{\Sigma}\vec{A}=\vec{\Sigma}\hat{\vec{r}}+{i\over ma}\beta K\left(\vec{\Sigma}\vec{p}\right)~.\end{equation}
As far as, $[\beta~,~K]=0$, it follows that $\left\{K~,~\beta
K\left(\vec{\Sigma}\vec{p}\right)\right\}=0$ and
$K\left(\vec{\Sigma}\vec{p}\right)$ can be used as a permissible
$K$-odd operator.

Our purpose is to construct such combination of $K$-odd operators,
which would be commuting with Dirac Hamiltonian. We can solve this
task by step by step. As a first trial expression let consider the
following operator

\[A_1=x_1\left(\vec{\Sigma}\hat{\vec{r}}\right)+ix_2K\left(\vec{\Sigma}\vec{p}\right)~.\]
Here the coefficients are chosen in such a way, that $A_1$ be
Hermitian, when $x_1~,~x_2$ are arbitrary real numbers. These
numbers must be determined from the requirement of commuting with
$H$. Let calculate

\[[A_1,H]=x_1\left[\left(\vec{\Sigma}\hat{\vec{r}}\right),H\right]
+ix_2\left[K\left(\vec{\Sigma}\vec{p}\right),H\right]~.\]
Appearing here commutators can be calculated easily. The result is
\[[A_1,H]=x_1{2i\over r}\beta K\gamma^5 - x_2{a\over
r^2}K\left(\vec{\Sigma}\vec{r}\right)~.\] One can see, that the
first term in right-hand side is antidiagonal, while the second
term is diagonal. So this expression never becomes vanishing for
ordinary real numbers $x_1~,~x_2$. Therefore we must perform the
second step: one has to include new odd structure, which appeared
on the right-hand  side of above expression. Hence, we are faced
to the new trial operator
\begin{equation}A_2=x_1\left(\vec{\Sigma}\hat{\vec{r}}\right)+ix_2K\left(\vec{\Sigma}\vec{p}\right)+ix_3K\gamma^5f(r)~.\end{equation}
Here $f(r)$ is an arbitrary scalar function to be determined. Let
calculate new commutator. We have \[[A_2,H]=x_1{2i\over r}\beta
K\gamma^5 - x_2{a\over r^2}K\left(\vec{\Sigma}\hat{\vec{r}}\right)
-x_3f'(r)K\left(\vec{\Sigma}\hat{\vec{r}}\right)-ix_32m\beta
K\gamma^5f(r)~.\] Grouping diagonal and antidiagonal matrices
separately and equating this expression to zero, we obtain
equation \[K\left(\vec{\Sigma}\hat{\vec{r}}\right)\left({a\over
r^2}x_2+x_3f'(r)\right) +2i\beta K\gamma^5\left({1\over
r}x_1-mx_3f(r)\right)=0~.\] This equation is to be satisfied, if
diagonal and antidiagonal terms become zero separately, i.e.,
\[{a\over r^2}x_2=-x_3f'(r)~~~~,~~~~{1\over
r}x_1=mx_3f(r)~.\] First of all, let us integrate the second
equation in the interval $(r~,~\infty)$. It follows
\[x_3f(r)=-{a\over r}x_2~.\]
Accounting this in the first equation, we have
\[x_2=-{1\over ma}x_1~.\]
Then we obtain also
\[x_3f(r)=-{1\over mr}x_1~.\]
Therefore finally we have derived the following operator  which
commutes with Dirac Hamiltonian
\begin{equation}\label{opcomDH}A_2=x_1\left\{\left(\vec{\Sigma}\hat{\vec{r}}\right)-{i\over ma}K\left(\vec{\Sigma}\vec{p}\right)+{i\over m}K\gamma^5{1\over r}\right\}~.\end{equation}
It is $K$-odd, in accord with above theorem.

If we turn to usual $\vec{\alpha}$ matrices using the relation
$\vec{\Sigma}=\gamma^5\vec{\alpha}$ and taking into account the
expression (\ref{Dirham}) of the Dirac Hamiltonian, $A_2$ can be
reduced to the more familiar form ($x_1$, as unessential common
factor, may be dropped)
\[A_2=\gamma^5\left\{\vec{\alpha}\hat{\vec{r}}-{i\over
ma}K\gamma^5\left(H-\beta m\right)\right\}~.\] This expression is
nothing but the Johnson-Lippmann operator that was introduced by
these authors in a very brief abstract in 1950. As to more
detailed settle, by our knowledge, it had not been published
neither then nor after.

It seems that our derivation of this operator is rather easy and
transparent. Moreover, in parallel, we have shown the
commutativity of this operator with the Dirac Hamiltonian.

In order to clear up its physical meaning, remark, that equation
(\ref{opcomDH}) may be rewritten in the form
\[A_2=\vec{\Sigma}\left(\hat{\vec{r}}-{i\over
2ma}\beta\left[\vec{p}\times\vec{l}-\vec{l}\times\vec{p}\right]\right)+{i\over
mr}K\gamma^5~.\] It is evident, that in nonrelativistic limit,
when $\beta\rightarrow 1$ and $\gamma^5\rightarrow 0$, this
operator reduces to the projection of LRL vector on the electron
spin direction, $A_2\rightarrow \vec{\Sigma}\vec{A}$ or because of
$\vec{l}\vec{A}=0$ it is a projection on the total $\vec{J}$
momentum direction.

After this it is clear that the Witten algebra can be derived by
identifying supercharges as \cite{KA}
\[Q_1=A~,~~~~Q_2=i{AK\over k}~.\]
It follows that
\[\{Q_1,Q_2\}=0~,~~\mbox{and}~~Q_1^2=Q_2^2=A^2~.\]
This last factor may be identified as a Witten Hamiltonian (N=2
supersymmetry).

As for spectrum, it is easy to show that the following relation
occurs \cite{KA}
\[A^2=1+\left({K\over a}\right)^2\left({H^2\over m^2}-1\right)~.\]
Because all operators entered here commute with each other, we can
replace them by their eigenvalues. Therefore we obtain energy
spectrum pure algebraically. In this respect it is worthwhile to
note full analogy with the classical mechanics, where closed
orbits were derived by calculating the square of the LRL vector
without solving the differential equations of motion \cite{Gol}.

In conclusion let underline once more that the degeneracy of
spectrum relative to interchange $k\rightarrow -k$, is connected
to the existence of conserved Johnson-Lippmann operator, which
takes its origin from the Laplace-Runge-Lenz vector. It is also
remarkable that the same symmetry is responsible for absence of
the Lamb-shift in this problem. Inclusion the Lamb-shift terms
into the Dirac Hamiltonian spoils commutativity of $A$ with $H$,
and consequently, above mentioned supersymmetry.

This work was suported  in part by Reintegration Grant No.
FEL.REG. 980767.

\end{document}